\documentclass[reprint, superscriptaddress, amsmath, amssymb, aps, pra]{revtex4-2}
\usepackage{graphicx}
\usepackage{dcolumn} 
\usepackage{xcolor}
\usepackage[absolute]{textpos}
\usepackage{amssymb}
\usepackage{bm}
\usepackage{braket}
\definecolor{darkblue}{rgb}{0,0.3,0.7}
\usepackage{hyperref, comment}
\usepackage{algorithm}
\usepackage{algpseudocode}
\usepackage{float}
\hypersetup{
colorlinks=true,
linkcolor=darkblue,
filecolor=blue,
citecolor=darkblue,  
urlcolor=darkblue,
}
\DeclareMathOperator{\Tr}{Tr}
\def\bx{\mathbf{x}}
\def\by{\mathbf{y}}
\def\bcD{{\bm{\mathcal{D}}}}

\def\cZ{\mathcal{Z}}
\def\cF{\mathcal{F}}
\def\cN{\mathcal{N}}

\usepackage[T1]{fontenc}     
\usepackage[utf8]{inputenc}  
\usepackage[scaled]{sourcecodepro} 

\begin{document}

\preprint{APS/123-QED}

\title{
Unorthodox parallelization for Bayesian quantum state estimation
}
\author{Hanson H. Nguyen}
\affiliation{School of Electrical, Computer, and Energy Engineering and Research Technology Office, Arizona State University, Tempe, Arizona 85287, USA}
\author{Kody~J.~H. Law}
\affiliation{School of Mathematics, University of Manchester, Manchester, M13 9PL, UK}
\author{Joseph~M. Lukens}
\email{jlukens@purdue.edu}
\affiliation{School of Electrical, Computer, and Energy Engineering and Research Technology Office, Arizona State University, Tempe, Arizona 85287, USA}
\affiliation{Elmore Family School of Electrical and Computer Engineering and Purdue Quantum Science and Engineering Institute, Purdue University, West Lafayette, Indiana 47907, USA}
\affiliation{Quantum Information Science Section, Oak Ridge National Laboratory, Oak Ridge, Tennessee 37831, USA}

\date{\today}

\begin{abstract}
Quantum state tomography (QST) allows for the reconstruction of quantum states through measurements and some inference technique under the assumption of repeated state preparations. Bayesian inference provides a promising platform to achieve both efficient QST and accurate uncertainty quantification, yet is generally plagued by the computational limitations associated with long Markov chains. In this work, we present a novel Bayesian QST approach that leverages modern distributed parallel computer architectures to efficiently sample a $D$-dimensional Hilbert space. Using a parallelized preconditioned Crank--Nicholson Metropolis-Hastings algorithm, we demonstrate our approach on simulated data and experimental results from IBM Quantum systems up to four qubits, showing significant speedups through parallelization. Although highly unorthodox in pooling independent Markov chains, our method proves remarkably practical, with validation \emph{ex post facto} via diagnostics like the intrachain autocorrelation time. We conclude by discussing scalability to higher-dimensional systems, offering a path toward efficient and accurate Bayesian characterization of large quantum systems.
\end{abstract}

\maketitle

\section{Introduction}
Quantum state tomography (QST) is a cornerstone of quantum information processing (QIP), allowing for the full characterization of quantum sources, gates, and channels~\cite{hradil1997quantum, Nielsen2000, james2001measurement,ALTEPETER2005105}. However, the number of parameters required to fully describe a quantum state scales exponentially with the number of qubits, which, while advantageous for developing complex information devices that surpass classical capabilities, poses significant challenges for characterizing large-scale quantum systems. In fact, the computational requirements of QST in large QIP systems far exceeds modern computational capabilities, thus rendering the characterization of larger QIP systems infeasible.

Although not circumventing fundamental scaling limits,
Bayesian inference is a promising paradigm for QST with robustness and measurement efficiency, pairing automatic uncertainty quantification with optimality guarantees in terms of minimum mean squared error---features that hold for any number and collection of measurements~\cite{Robert1999, MacKay2003, blume2010optimal}. In Bayesian inference, a prior distribution is updated upon obtaining new evidence. In the context of QST, the probability distribution is a complex and high-dimensional function of parameters that describe the density matrix $\rho$, and new evidence is simply more measurements.
In lieu of analytical evaluation (impossible for all but the simplest models), numerical techniques such as Markov chain Monte Carlo (MCMC) \cite{Metropolis1953,Hastings1970,Cotter2013} and sequential Monte Carlo~\cite{delMoral2006,Chopin2020,Dai2022} are often used to sample from the full probability distribution, resulting in a built-in method of characterizing uncertainty on a Bayesian mean estimate of $\rho$. However, despite a variety of Monte Carlo algorithms proposed in QIP~\cite{Seah2015, Granade2016, Granade2017, Mai2017, Williams2017, Lukens_2020},
efficient Bayesian QST is still met with the challenge of poor sampling efficiency due to the complex high-dimensional probability distribution, leading to slow convergence times.

In this work, we propose a Bayesian QST technique that leverages parallelization through noncommunicating processing cores to significantly reduce wall clock time for Bayesian QST. Our method relies on independent preconditioned Crank--Nicholson (pCN)~\cite{Cotter2013} Metropolis--Hastings~\cite{Metropolis1953,Hastings1970} chains that combine to produce samples from the Hilbert space of a given $\rho$. 
Previously applied only serially in QST~\cite{Lukens_2020,Chapman:22,Lu2022b}, our parallelized version directly pools all samples from independent chains---a highly unorthodox use of MCMC trading its asymptotic ``many-iteration'' guarantees for a ``many-chain'' scaling amenable to parallelization yet susceptible to bias.

Building on recent developments in statistical inference that leverage independent chains for parallelization and validate their legitimacy through \emph{ex post facto} metrics~\cite{Chen2016, Heng2023}, we embrace such unconventional parallel chains and confirm their practical ability to estimate the Bayesian mean density matrix with faster speeds than serial alternatives.
In doing so, we demonstrate a QST scheme that is both computationally efficient and robust for reconstructing higher-dimensional QIP systems. After describing the algorithm, we show inference results on both simulated and experimental datasets from up to four qubits, finding significant computational speedups, achieving $\gtrsim$100-fold lower errors for the same average wall clock time in 1024-chain examples. Finally, we conclude with directions for future research in parallelizable Bayesian QST, including an outlook on expanding to even higher-dimensional QIP systems.   

\begin{figure*}[!t]\centering
\includegraphics[width=\textwidth]{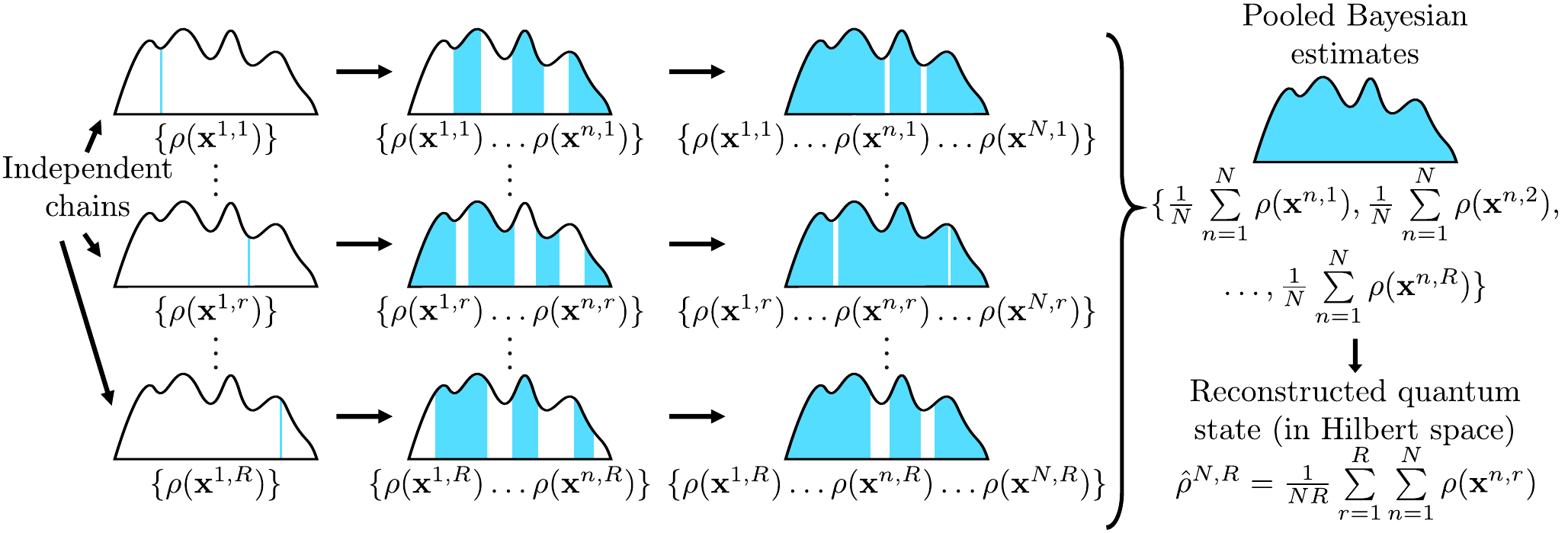}
    \caption{Visualization of parallel pCN, where density matrix samples from multiple independent chains explore the full Hilbert space more accurately than each individual chain.}
    \label{diagram}
\end{figure*}

\section{Background}
\label{sec:background}
The density matrix $\rho$ for a system of $Q$ qubits is a $D \times D$ matrix, where $D = 2^Q$ is the dimension for the Hilbert space of the entire system. The density matrix $\rho$ represents a quantum state in this complex $D$-dimensional Hilbert space; it is positive semidefinite  ($\bra{\psi}\rho\ket{\psi} \geq 0$ for any $\ket{\psi}\in\mathbb{C}^{D\times 1}$) with unit trace ($\Tr\rho = 1$) and can be written as
\begin{equation}
\label{eq:rho}
\rho = \sum_{i=1}^D\sum_{j=1}^D a_{ij}\ket{i}\bra{j},
\end{equation}
where $\{\ket{i}\}$ are basis vectors of the Hilbert space. With these constraints, $\rho$ has $D^2-1$ degrees of freedom, and thus a minimum of $D^2-1$ independent real parameters are required for its description~\cite{ALTEPETER2005105}. 

To recreate $\rho$, measurements in various bases $k$ can be performed, each of which can be described by a positive operator-valued measure (POVM) consisting of a collection of operators $\Lambda_{kl}$, each connected to an outcome probability $p_{kl} = \Tr{\Lambda_{kl}\rho}$~\cite{Nielsen2000}, for outcome $l$. Then some inference method collects the results from all bases and seeks to reconstruct $\rho$. At the informationally complete extreme, linear inversion tomography generally considers $4^{Q}-1$ measurements for a $Q$-qubit system, reconstructing $\rho$ by solving the set of linear equations that describe the relationship between outcome probabilities and the unknown parameters~\cite{Nielsen2000}. At the other extreme of minimal measurements, classical shadow tomography focuses on estimating observables and not the quantum state \emph{per se} by associating each measurement with a ``classical shadow''~\cite{Huang2020}. In maximum likelihood estimation (MLE)---arguably the most common approach in QIP---a constrained optimization of $\rho$ based on measurement outcomes returns the $\rho$ associated with the highest probability of producing the observations~\cite{hradil1997quantum,james2001measurement}. Finally, in Bayesian QST, Bayes' theorem is used to update a prior distribution on the density matrix $\rho$ based on  measurement results~\cite{blume2010optimal}.
Incidentally, neither MLE nor Bayesian inference places formal requirements on the quantity or quality of measurements chosen; each can obtain an answer from any measurement set, irrespective of questions of informational completeness. Yet Bayesian inference provides a safety net, in that it also furnishes uncertainty commensurate with the measurements chosen and therefore signals against unjustified conclusions from incomplete data.

To further elaborate on the Bayesian inference approach, consider $\rho(\bx)$ where $\bm{\mathrm{x}}$ is the set of parameters describing $\rho$. With parameters $\bm{\mathrm{x}}$, Bayes' theorem is expressed in the convenient form
\begin{equation}
\label{eq:Bayes}
\pi(\bx) = \frac{1}{\cZ} L_\bcD(\bx) \pi_0(\bx)
\end{equation}
where $\pi(\bx)= \Pr(\bx|\bcD)$ is the posterior of interest, $L_\bcD(\bx) \propto \Pr(\bcD|\bx)$ is the likelihood over the observed data $\bcD$, $\pi_0(\bx) = \Pr(\bx)$ denotes the prior, and $\mathcal{Z} = \int d\bx\,L_\bcD(\bx) \pi_0(\bx)$ is a normalizing constant, which does not need to be computed in our approach.

Performing integrals over Eq.~(\ref{eq:Bayes}) is computationally intractable for nearly all problems of interest (particularly in QST). MCMC algorithms can be implemented to sample from $\pi(\bx)$ and thereby approximate any quantity of interest using a sequence of $N$ discrete samples $\bx^{1:N} = \{\bm{\mathrm{x}}^{}, \bm{\mathrm{x}}^{}, \dots, \bm{\mathrm{x}}^{}\}$. From these samples, any quantity of interest can be estimated; in this study, we focus specifically on the Bayesian mean density matrix $\braket{\rho} = \int d\bx\,\pi(\bx)\rho(\bx)$---the minimum mean-squared error estimator~\cite{Robert1999} or $\rho$ which we can approximate by the $N$-sample mean:
\begin{equation}
\label{eq:BME1}
\hat{\rho}^N = \frac{1}{N}\sum_{n=1}^N \rho(\bx^n)
\end{equation} 
Modifications to traditional Metropolis--Hastings algorithms for obtaining the samples $\bx^{1:N}$ have often focused on improving the step-size/acceptance-rate tradeoff. For instance, pCN implements a proposal scheme that promotes controlled changes in the sample: $\bm{\mathrm{x}}' = \sqrt{1-\beta^2}\bm{\mathrm{x}} + \beta\bm{\xi}'$ to overcome inefficient sampling of high-dimensional systems and improve convergence times \cite{Cotter2013,Lukens_2020}. In our scheme, we also implement parallelization to further reduce convergence times by encouraging random independent explorations of the Hilbert space. Running $R$ independent pCN chains and merging the $N$ samples from each into $\bx^{1:N,1:R} = \{\bx^{1,1},\bx^{2,1},\dots,\bx^{N,R}\}$, we can replace $\hat{\rho}^N$ with the $R$-chain estimate
\begin{equation}
\label{eq:BME2}
\hat{\rho}^{N,R} = \frac{1}{NR}\sum_{r=1}^R \sum_{n=1}^N \rho(\bx^{n,r})
\end{equation}
Given sufficient computational resources, the limit $R\rightarrow\infty$ can be obtained with no increase in the length of an individual chain; therefore as long as accuracy is improving with $R$, we can increase the number of samples without extending the overall wall clock time, potentially leading to massively improved efficiency.

However, it is important to note that our parallelization approach is unconventional, as MCMC is inherently iterative and serial. In other words, the limit 
\begin{equation}
\label{limN}
\lim_{N\to\infty}\hat{\rho}^N = \braket{\rho}
\end{equation}
known as consistency, on which the whole program of MCMC is based, does \emph{not} hold for $R\to\infty$; i.e., the estimator $\hat{\rho}^{N,R}$ is asymptotically biased in general:
\begin{equation}
\label{limR}
\lim_{R\to\infty}\hat{\rho}^{N,R} \neq \braket{\rho}.
\end{equation}

While algorithms such as coupled chains~\cite{Jacob2020,Heng2023} and parallelized sequential Monte Carlo (SMC) \cite{Dai2022,Liang2024} offer routes such that Eq.~(\ref{limR}) does indeed become an equality, they are considerably more complex and computationally demanding than simply pooling independent MCMC chains, which we explore here, finding  unorthodox improvements in convergence time and exploration of the Hilbert space.

Figure~\ref{diagram} presents a heuristic sketch of our method in the ideal case. Prior to convergence, each chain independently explores different regions of the Hilbert space through random sampling. Pooling these regions accelerates convergence by leveraging the diverse initial sampling areas and random steps of each chain and averaging across them, ensuring coverage of the full posterior distribution (represented here as a one-dimensional function for visual simplicity), which may not be explored by a single chain alone. Provided the impacts of bias are low---unproven theoretically but tested empirically below---this simple method enhances both speed and robustness through such independent and parallelizable chains.

\section{Proposed Approach}
\label{sec:algorithm}
The single-chain workflow that we parallelize here builds on current Bayesian QST methods, specifically (i)~pCN MCMC sampling---applied to QST first in 2020~\cite{Lukens_2020}---and (ii)~Bures-based parametrization~\cite{AlOsipov2010}---combined with pCN-based QST in 2021~\cite{Lohani2021}. Although detailed in prior work, we review both algorithms briefly here for reference.
The latter, described in Algorithm~\ref{a1}, generates a density matrix from parameters that are independent and identically distributed (i.i.d.) according to the standard normal distribution $\cN(0,1)$; the result is a density matrix $\rho$ drawn from the Bures metric, salient in quantum mechanics as the only monotone, Fisher-adjusted, and Fubini--Study-adjusted density matrix distribution~\cite{Sommers2003}. Algorithm~\ref{a1} requires $4D^2$ parameters, $\sim$4-fold overparametrization beyond the minimum of $D^2-1$ for an arbitrary density matrix~\cite{ALTEPETER2005105}, but is extremely efficient computationally, automatically enforcing both positive semidefiniteness and the desired Bures prior with standard normal parameters.

Algorithm~\ref{a2} details , which is parallelized and pooled in our method. The pCN addition ideally reduces the standard tradeoff for random-walk Metropolis--Hastings in which large step size and high  acceptance rate become increasingly difficult to satisfy as dimensions increase~\cite{Cotter2013}. In addition to the  basic pCN procedure (Lines 3--12), Algorithm \ref{a2} explicitly highlights our approach to step-size adaptation (Lines 13--21), in which we utilize counter $A$ to check the acceptance rate every $M_A$ steps ($M_A=500$ in our simulations), increasing or decreasing $\beta$ in order to keep $A/M_A\in[0.2,0.6]$---an effective range for mixing with pCN~\cite{Cotter2013,law2014proposals}. Lines 22--24 show how thinning $T$ is implemented; even though the total chain is $NT$ points long, we store $N$ samples only to maintain manageable file sizes.
We have found both procedures extremely useful in implementing Bayesian QST in practice.

Finally, while we focus on a standard normal prior (in line with the Bures distribution) and multinomial likelihood [Eq.~(\ref{likelihood})], other priors and likelihoods can be handled easily, with the Algorithm~\ref{a2}'s only modifications occurring in Steps 3--5; we point the reader to Ref.~\cite{Lukens2021c} for a listing of pCN proposals for other typical priors and Refs.~\cite{Lu2022b,Chapman:22} for example non-multinomial likelihoods tailored to specific experimental conditions.

\subsection{Steps}
Our complete parallelized Bayesian QST approach proceeds as follows.

\begin{enumerate}
    \item  Measure $D\times D$ matrix $\rho$, the unknown quantum state, $P$ times each in $K$ positive operator-valued measures $\{\bm{\Lambda}_k\}_{k\in\{1,...,K\}}$, each with $D$ outcomes, defined as $\bm{\Lambda}_k=\{\Lambda_{k1},...,\Lambda_{kD}\}$. Store all $KD$ measurement outcomes  as $\bcD=\{c_{11},c_{12}...,c_{KD}\}$. (Note that $\sum_{k,l} c_{kl} = KP$.) 
    
    \item  Parameterize $\rho(\bx)$ with length-$4D^2$ real vector $\bx$ according to Steps 2--6 of Algorithm~\ref{a1}. 

    \item Define prior \begin{equation}
    \pi_0(\bx) =  \prod_{k=1}^{4D^2} (2\pi)^{-1/2} e^{-x_k^2/2}, 
    \end{equation}
    i.e., $\bx\sim\cN(0,I_{4D^2})$, where $I_{4D^2}$ is the $4D^2$$\times$$4D^2$ identity matrix.
    
    \item  Define likelihood \begin{equation}
    \label{likelihood}
    L_\bcD(\bx) = \prod_{k=1}^K \prod_{l=1}^D \left[\Tr \Lambda_{kl}\rho(\bx)\right]^{c_{kl}}
    \end{equation}
    \item For each chain $r$ or $R$ independent chains, draw and store $N$ samples $\bx^{1:N,r}$ according to Algorithm~\ref{a2}.
    \item Use the resultant samples to estimate expected values of any function $\phi(\rho)$ as \begin{equation}
    \left\langle\phi(\rho)\right\rangle \approx \hat{\phi}^{N,R} = \frac{1}{NR}\sum_{r=1}^R\sum_{n=1}^N\phi(\rho(\bm{\mathrm{x}}^{n,r})),
    \end{equation}
    where we use the superscript notation $N,R$ to denote an estimate of the Bayesian mean from $R$ chains with $N$ samples each.
\end{enumerate}

\begin{algorithm}[H]
\caption{Drawing a density matrix from the Bures distribution}\label{a1}
\begin{algorithmic}[1]
    \State Draw length-$4D^2$ vector $\bx\sim~\cN(0,I_{4D^2})$.
    \State Define $D\times D$ matrix $G$ with real components populated by the first $D^2$ components of $\bx_{1:D^2}$ and imaginary components populated by $\bx_{D^2+1:2D^2}$.
    \State Define $D\times D$ matrix $H$ with real elements $\bx_{2D^2+1:3D^2}$ and imaginary components $\bx_{3D^2+1:4D^2}$. 
    \State Perform QR decomposition to obtain $ = R$, where $Q$ is unitary and $R$ is upper triangular.
        \State Compute $U = Q\Lambda$, where $\Lambda =  
        \begin{pmatrix} 
        \frac{r_{11}}{|r_{11}|} & \cdots& 0 \\
         \vdots & \ddots & \vdots \\
         0 & \cdots & \frac{r_{DD}}{|r_{DD}|} 
        \end{pmatrix}$~\cite{Mezzadri2006}.
    \State Compute $W = (U + I_{D^2})G$ and define $\rho(\bx) = \frac{WW^\dagger}{\Tr{WW^\dagger}}$~\cite{AlOsipov2010}.
\end{algorithmic}
\end{algorithm}
\begin{algorithm}[H]
\caption{pCN for single chain $r$}\label{a2}

\begin{algorithmic}[1]
    \State Set $A=0$ and $\beta=0.1$ and draw $\by^{0, r} \sim \cN(0,I_{4D^2})$.
    \For{$j=1,...,NT$}
    \State Draw $\bm{\eta} \sim \mathcal{N}(0,I_{4D^2})$.
    \State Propose  $\by' = \sqrt{1-\beta^2}\by^{j-1, r}+\beta\bm{\eta}$
    \State Construct $\rho(\by^{j-1, r})$ and $\rho(\by')$ through Steps 2--6 of Algorithm~\ref{a1} and compute $L_\bcD(\by^{j-1,r})$ and $L_\bcD(\by')$ from Eq.~\ref{likelihood}.
    \State Draw $\alpha\sim \mathbb{U}(0,1)$. 
    \If{$\log\alpha < \log L_\bcD(\by')-\log L_\bcD(\by^{j-1,r})$}
    \State $\bm{\mathrm{y}}^{j, r} \leftarrow \by'$. 
    \State $A\leftarrow A+1$.
    \Else{}
    \State $\by^{j, r} \leftarrow \by^{j-1, r}$ 
    \EndIf    
    \If{$j = 0 \mod M_a$}
    \If{$A/M_A > 0.6$}
    \State $\beta \leftarrow 1.1\beta$.
    \EndIf
    \If{$A/M_A < 0.2$}
    \State $\beta \leftarrow \beta/1.1$.
    \EndIf
    \State $A\leftarrow 0$.
    \EndIf
    \If{$j=0\mod T$}
    \State $\bx^{j/T,r} \leftarrow \by^{j,r}$.
    \EndIf
    \EndFor
    \State Output $N$ samples $\bx^{1:N,r} = \{\bx^{1,r},...,\bx^{N,r}\}$.
\end{algorithmic}
\end{algorithm}

\section{Simulated Results}
\label{sec:simulation}
To test the effective scaling of parallel chains, we first implement our algorithm on simulated results from up to four qubits. For a ground truth, we choose random states from the Bures distribution (leveraging Algorithm~\ref{a1}), specifically one such state for each qubit number $Q\in\{1,2,3,4\}$. To select the number of state copies per measurement setting (which we take as all possible Pauli measurement combinations, i.e., $K=3^Q$), we aim for fidelity $F \approx0.96$ with respect to the ground truth in our algorithm, which we found could be achieved with $P = 25\cdot 2^{Q}$ shots, for a total of $KPD=25\cdot 6^Q$ state copies in each simulated experiment. (Note that the number of state copies does not change any aspect of the algorithm itself; we simply seek to operate in a regime where the Bayesian mean is relatively close to the ground truth---the desired scenario in practice.) Each pCN chain $r$ is $NT$ steps long, where, we save a fixed $N=2^{10}$ samples $\bx^{1:N,r}$ but increase the thinning factor $T$ to improve mixing. All tests are completed in 64-bit MATLAB on a 2.6~GHz computer running 48 cores.

The individual chain performance is critical to the final pooling step, since strongly correlated chains will not scale favorably with $R$. To assess mixing, we consider the autocorrelation function (ACF) defined as
\begin{equation}
\label{eq:ACFrho}
c[l] = \frac{1}{C}\sum_{n=1}^{N-l_{\max}} \mathfrak{Re}\left\{  \Tr [\rho(\bx^n) - \hat{\rho}^N]^\dagger [\rho(\bx^{n+l}) - \hat{\rho}^N]\right\}
\end{equation}
where $C$ is a normalization constant such that $c[0]=1$. This definition can be viewed as a scalar summary of the individual components [Eq.~(\ref{eq:rho})]. Expanding Eq.~(\ref{eq:ACFrho}) gives
\begin{multline}
\label{eq:ACFa}
c[l] = \frac{1}{C} \sum_{i=1}^D\sum_{j=1}^D\sum_{n=1}^{N-l_{\max}}\Big\{ \mathfrak{Re}\left[a_{ij}^n-\hat{a}_{ij}^N\right] \mathfrak{Re}\left[a_{ij}^{n+l}-\hat{a}_{ij}^N \right] \\
+ \mathfrak{Im}\left[a_{ij}^n-\hat{a}_{ij}^N\right] \mathfrak{Im}\left[a_{ij}^{n+l}-\hat{a}_{ij}^N \right] \Big\},
\end{multline}
with the definitions $a_{ij}^n=\braket{i|\rho(\bx^n)|j}$ and  $\hat{a}_{ij}^N=\braket{i|\hat{\rho}^N|j}$. In other words, $c[l]$ averages the correlation functions of all real and imaginary components separately, making it a useful---albeit relatively coarse---overview of the pCN mixing properties. (As an aside, we note that by defining the ACF in terms of $\rho$ rather than the underlying parameters $\bx$ avoids any potential issues with nonidentifiability, i.e., the fact multiple parameter sets $\bx$ can produce the same $\rho$.)

\begin{figure}[tb!]\includegraphics[width=\columnwidth]{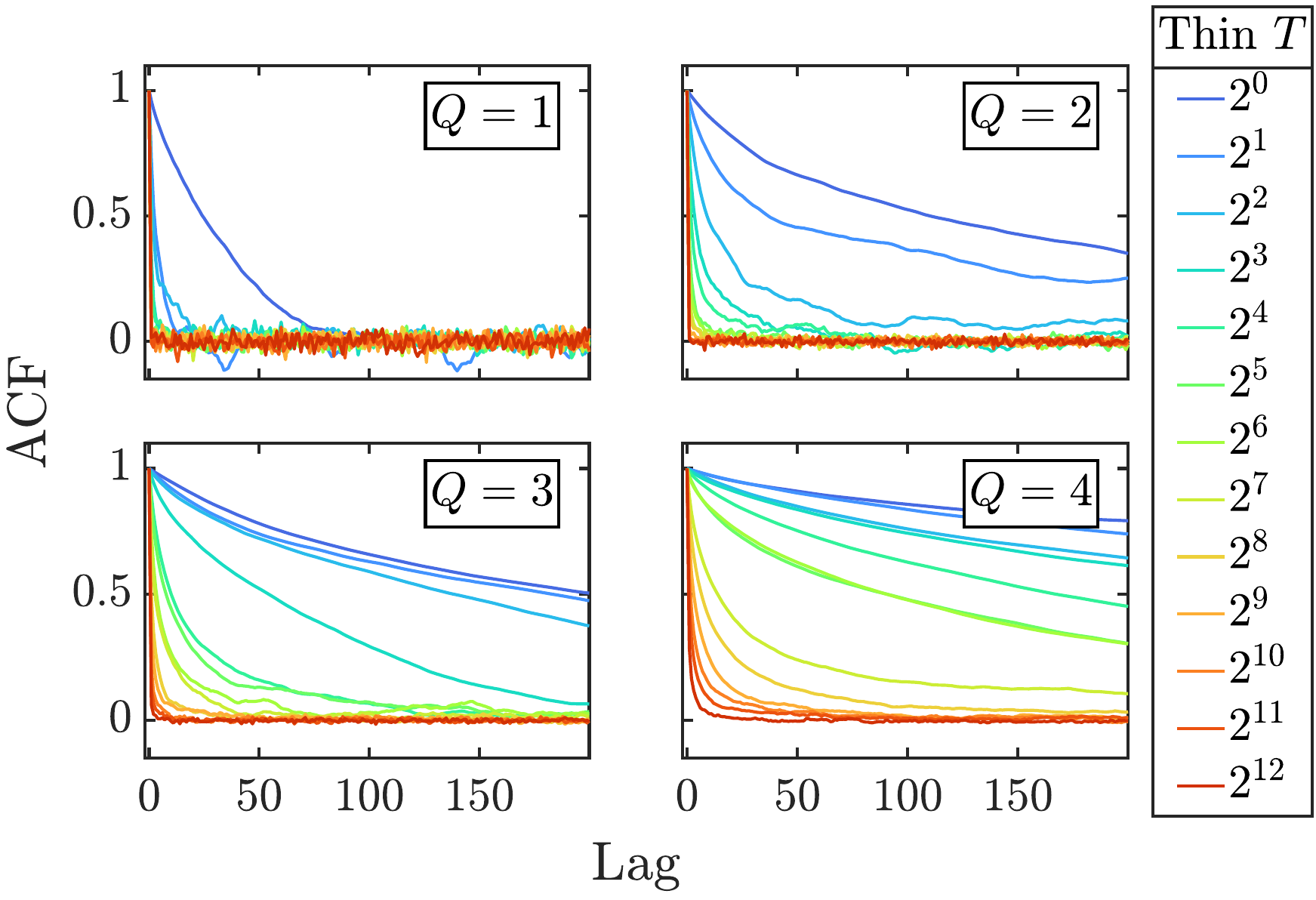}
    \caption{Autocorrelation functions (ACFs) for one randomly chosen pCN chain for each $Q\in\{1, 2, 3, 4\}$ qubits. Each line represents a different thinning value $T$, and $l_{\max} = 200$ is the maximum lag for all cases.}
    \label{fig:ACF}
\end{figure}

The ideal case of fully independent samples would lead to the delta function $\mathrm{ACF}=\delta[l]$. As shown in Fig.~\ref{fig:ACF}, our chains indeed approach this limit as $T$ (and thus the chain length $NT$) increases, with the lower qubit numbers doing so with shorter chains than higher qubits. By considering an integrated autocorrelation time (IACT) defined as~\cite{Chopin2020}
\begin{equation}
\label{eq:IACT}
\tau = 1 + 2\sum_{l=1}^{l_\mathrm{max}} c[l],
\end{equation}
this mixing length can be quantified. For example, at the highest thinning $T=2^{12}$ we compute $\tau \in \{2.65, 2.60, 2.61, 4.09\}$ for $Q\in\{1,2,3,4\}$, respectively, in Fig.~\ref{fig:ACF}.

Having analyzed serial mixing for individual chains, we next explore scaling with  parallelization. For each qubit number $Q$ and thinning $T$, we pool $R$ independent chains into the estimate $\hat{\rho}^{N,R}$ [Eq.~(\ref{eq:BME2})] which we then compare against the true Bayesian mean $\braket{\rho}$ through the Frobenius error 
\begin{equation}
\label{eq:fro}
\epsilon_F^2 = \left\| \hat{\rho}^{N,R} - \braket{\rho} \right\|_{F}^2 \approx \left\| \hat{\rho}^{N,R} - \hat{\rho}_{T_\mathrm{max}}^{N,R_\mathrm{max}} \right\|_{F}^2, 
\end{equation}
where we have approximated the true mean $\braket{\rho}$ with the estimator $\hat{\rho}_{T_\mathrm{max}}^{N,R_\mathrm{max}}$ corresponding to the maximum thinning $T_\mathrm{max}$ and number of chains $R_\mathrm{max}$ (since we do not have independent access to $\braket{\rho}$ directly). We focus on error with respect to $\braket{\rho}$ rather than the ground truth since these are in general different, and the mark of a successful Bayesian algorithm is to approximate the posterior target $\pi(\bx)$ (and hence quantities like $\braket{\rho}$ calculated therefrom). Defined in this way, $\epsilon_F^2$ should therefore approach zero regardless of the number of measurements in $\bcD$.

\begin{figure}[b!]\includegraphics[width=\columnwidth]{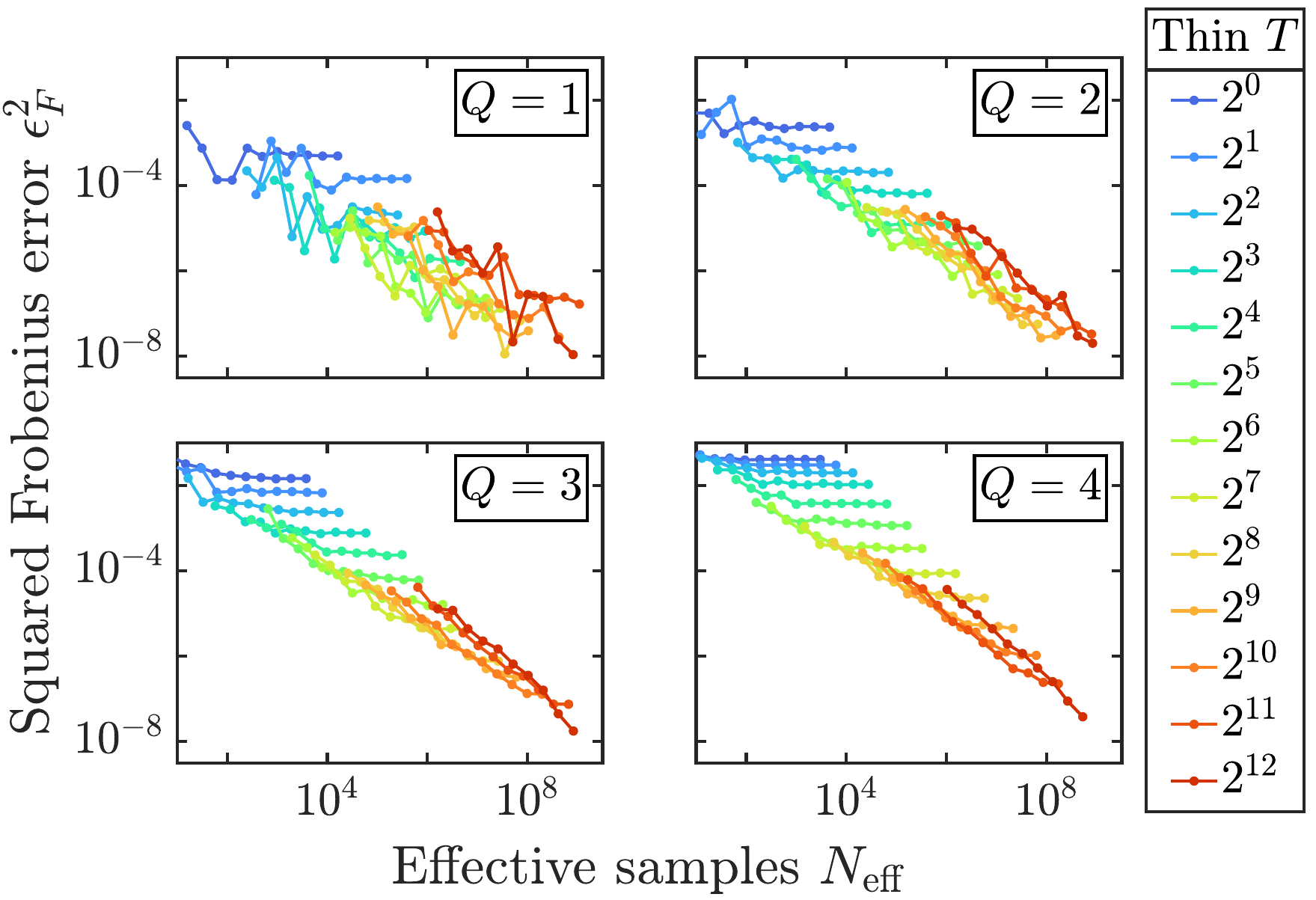}
    \caption{$\epsilon_F^2$ scaling for $Q\in\{1,2,3,4\}$ qubits. Each line represents constant time at a thinning $T$.}
    \label{fig:frob}
\end{figure}

Figure~\ref{fig:frob} plots $\epsilon_F^2$ against the effective number of samples defined using the IACT as $N_\mathrm{eff} = NR/\tau$. Each curve corresponds to a specific $Q$ and $T$, with $R\in[1,R_\mathrm{max}]$. Consequently, we can view each as a contour of constant wall clock time. Ideal parallel scaling corresponds to $\epsilon_F^2 \sim \mathcal{O}(1/N_\mathrm{eff})$. In reality, we find scaling more like $\epsilon_F^2 \sim A + B/N_\mathrm{eff}$: $\epsilon_F^2\sim\mathcal{O}(1/N_\mathrm{eff})$ holds initially until leveling off with some residual bias $A$. Unsurprisingly, this bias plateau depends on the serial chain length $NT$, highlighting the inherent trade-off for the parallelization method: one would like to keep $T$ as small as possible to minimize wall clock time; however, reducing $T$ leads to biases that, depending on the desired accuracy, could preclude sufficient convergence to $\braket{\rho}$.

\begin{figure}[tb!]\includegraphics[width=\columnwidth]{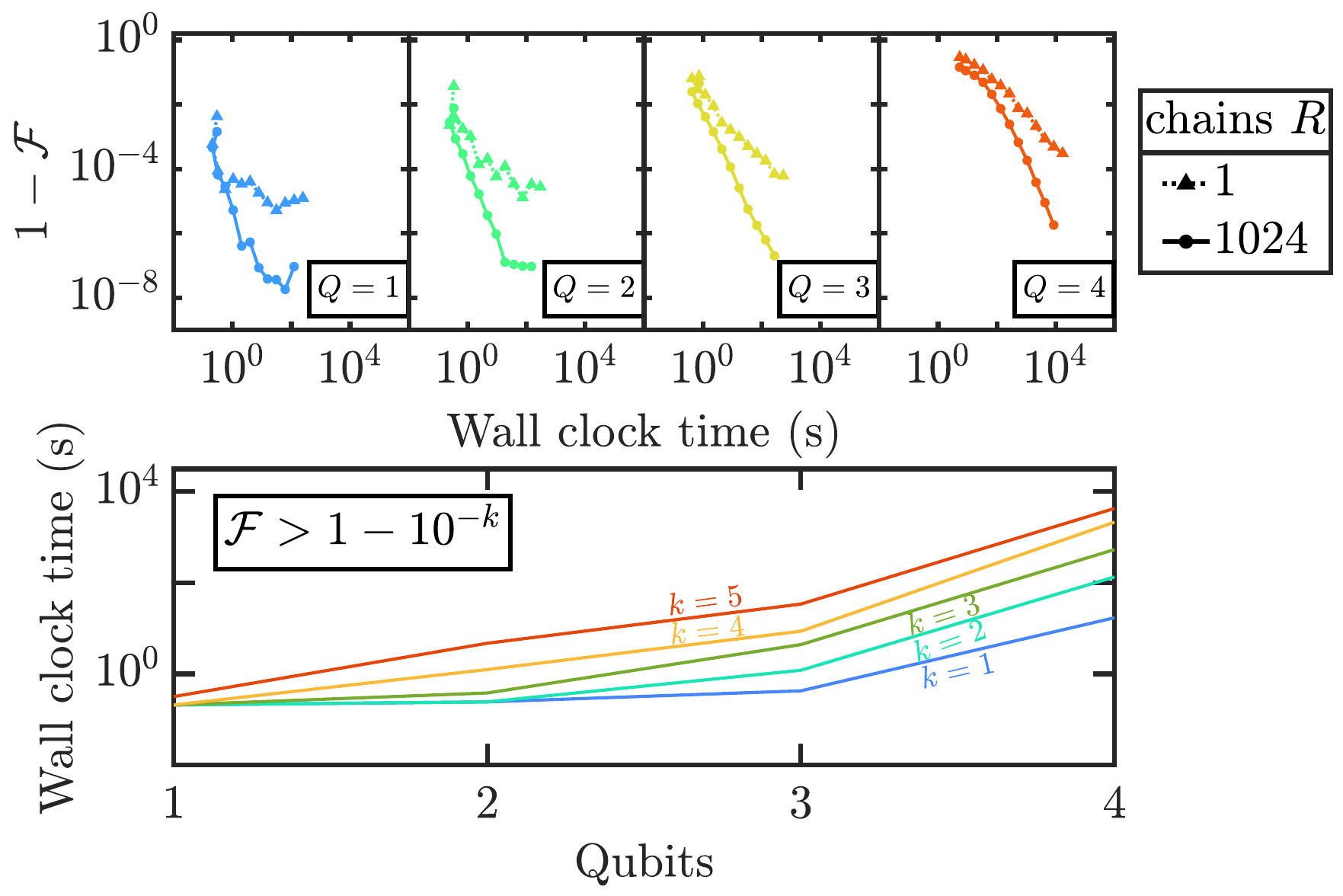}
    \caption{Wall clock time analysis on simulated datasets. (a)~Fidelity with respect to Bayesian mean as a function of time, for 1024 independent chains and one single chain. (b) Observed wall clock time required to reach specific fidelity thresholds $\cF>1-10^{-k}$ for the $R=1024$-chain cases.
    }
    \label{fig:time}
\end{figure}

Notwithstanding any such tradeoff questions, Fig.~\ref{fig:frob} highlights the clear error-reducing advantages of parallelization, particularly for examples with $T\gtrsim2^7$. Indeed, the ultimate success of any parallelization method centers on its ability to achieve a given accuracy in less time than a serial competitor. Figure~\ref{fig:time}(a) compares these times for both serial ($R=1$) and parallel ($R=1024$) pCN, where we use the fidelity 
\begin{equation}
\label{eq:fid}
\cF = \left(\Tr\sqrt{\sqrt{\braket{\rho}} \hat{\rho}^{N,R} \sqrt{\braket{\rho}}} \right)^2
\end{equation}
in place of Frobenius error $\epsilon_F^2$, in view of its greater ubiquity in QIP. 
Since our tests are run with 48 workers on a desktop computer, rather than a fully parallelized resource with $R\geq1024$ cores, we define the average wall clock time as 
$48/1024$ of the total run time, in order to correct for the limited number of cores available in our tests---a specific definition useful for estimating the equivalent fully parallelized absolute times, though not impacting the overall scaling trends.
In all cases,  $R=1024$ independent chains attain error $1-\cF$ at least two orders of magnitude lower than a single chain at the same wall clock time, at the highest serial chain length $NT$ (the rightmost points on each curve).

As another informative way to express the results in Fig.~\ref{fig:time}(a), we determine the minimum average wall clock time at which a particular fidelity threshold  $\cF>1-10^{-k}$ ($k\in\{1,2,...,5\}$) is reached. Plots of these findings against qubit number $Q$ in Fig.~\ref{fig:time}(b) reveal the natural tradeoff between time and accuracy at the available number of cores ($R_{\max}=1024$)---a tradeoff that should become increasingly less of an issue as $R\rightarrow\infty$. Nonetheless, our results as-is suggest the feasibility of pushing to larger Hilbert spaces with reasonable resources (time and memory).

\section{Experimental Results}
\label{sec:experiment}

\begin{figure}[t!]\includegraphics[width=\columnwidth]{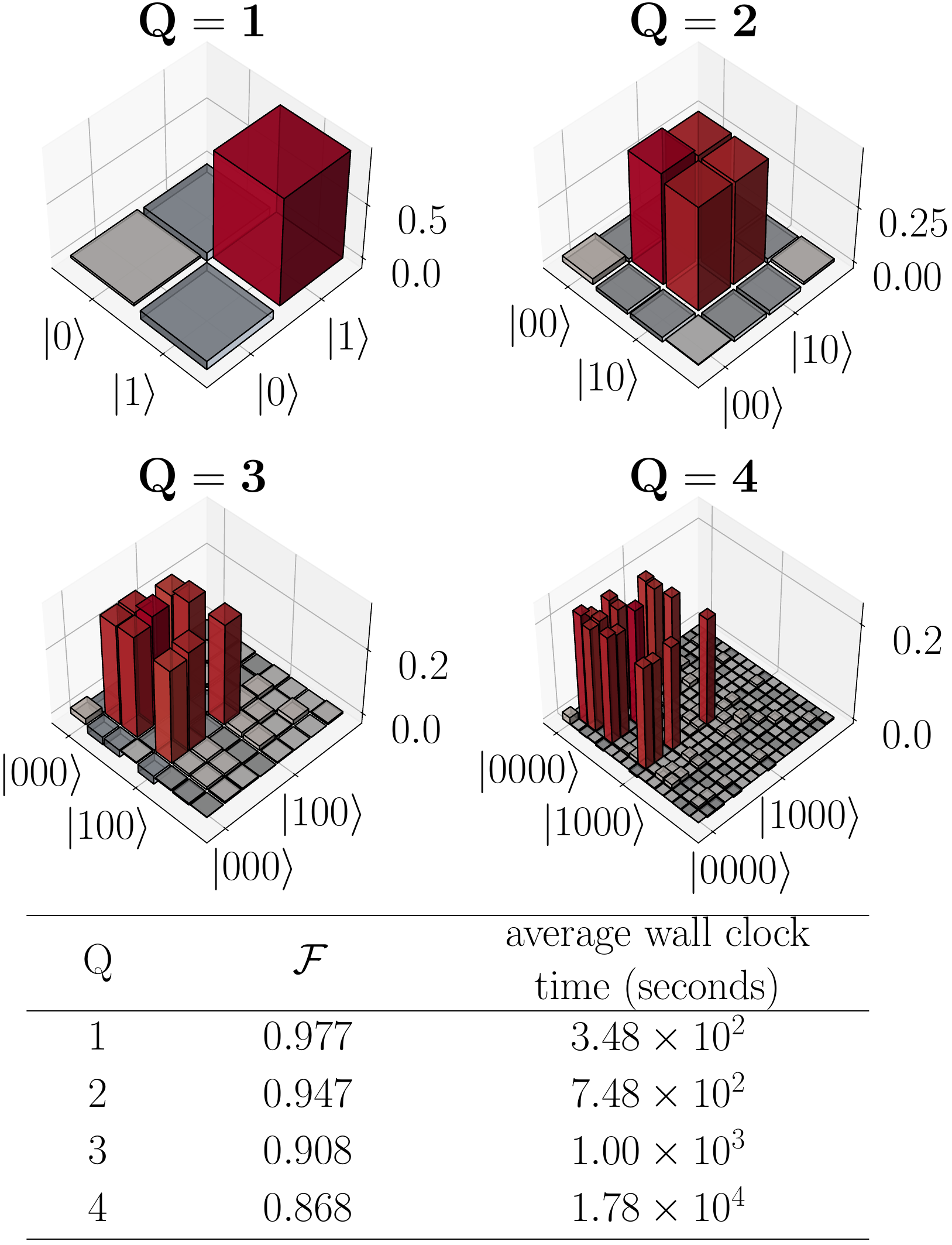}
    \caption{Real parts of reconstructed W states with parallel chains for $Q\in\{1,2,3,4\}$ qubits and $R=1024$ with respective fidelities ($\cF$) and wall clock time. All imaginary components (not shown) have moduli less than 0.04.}
    \label{fig:density matrices}
\end{figure}

In order to test the effectiveness of the parallelization technique on experimental systems, we perform measurements on the IBM Quantum processor Kyoto (\texttt{ibmq\_kyoto})~\cite{gadi_aleks_2019, qiskit2024} for quantum circuits designed to generate ideal W states, defined for $Q \in \{ 1, 2, 3, 4\}$ qubits as 
\begin{equation}
\label{eq:W}
\ket{W_{Q}} = \frac{1}{\sqrt{Q}}\sum_{q=1}^{Q}\ket{0}^{\otimes (q-1)}\ket{1}\ket{0}^{\otimes (Q - q)}.
\end{equation}
The same measurement protocols from Sec.~\ref{sec:algorithm} are used in these experiments, i.e., all $3^Q$ Pauli configurations with $P= 25\cdot 2^{Q}$ shots each. Since our focus here is on applying---rather than analyzing---the method, we report results at the highest thinning ($T=2^{12}$) and chain number ($R=2^{10}$) and estimate overlap  with respect to the target W state through the fidelity $\cF=\braket{W_Q|\hat{\rho}^{N,R}|W_Q}$. Note the difference in this definition compared to Eq.~(\ref{eq:fid}), which is designed to compare the point estimator $\hat{\rho}^{N,R}$ with the Bayesian mean; in contrast,  now we consider the mean of the observable $\ket{W_Q}\bra{W_Q}$---the target state, but not necessarily the ground truth under experimental noise.

Figure~\ref{fig:density matrices} plots the Bayesian mean estimate $\hat{\rho}^{N,R}$, sample fidelity, and wall clock time for each $Q$. Visually each result shows good agreement with an ideal W state, with fidelity reducing with dimension, unsurprising under the assumption of random errors: $\cF_{Q=1} = 0.977$, $\cF_{Q=2} = 0.947$, $\cF_{Q=3} = 0.908$, and $\cF_{Q=4} = 0.868$. 
In lieu of the ground truth (which is not directly available in experiment), we also perform MLE with the Qiskit Experiments 0.7.0 library \cite{Kanazawa2023}, finding: $\cF_{Q=1} = 0.992$, $\cF_{Q=2} = 0.971$, $\cF_{Q=3} = 0.935$, and $\cF_{Q=4} = 0.904$---close to the Bayesian estimates and thereby suggesting no major issues in the formalism. In fact, the slightly lower fidelities in the Bayesian case are consistent with its relative caution compared to MLE in returning low-rank estimators~\cite{blume2010optimal}.

\section{Discussion}
\label{sec:discussion}
Our unconventional approach offers valuable insight into accelerating Bayesian QST through parallelization. 
While not attaining ideal strong parallel scaling of the form $\epsilon_F^2 \sim \mathcal{O}(1/R)$, our simple method nevertheless demonstrates remarkable \emph{practical} gains in Bayesian QST, as evidenced by the observed $\sim$100-fold reductions in $1-\cF$ for the same wall clock time for $R=1024$ independent chains compared to just $R=1$ [Fig.~\ref{fig:time}(a)]. Scaling with qubit number suggests continued expansions to larger Hilbert spaces; 
although highly tenuous to extrapolate, rough fitting of Fig.~\ref{fig:time}(b) suggests that full Bayesian QST of a six-qubit system with $\cF>0.99$ may be possible with a wall clock time of just $2.34\times 10^4$~s---compared to the near week-long time for Bayesian QST on the same desktop machine and Hilbert space dimension in Ref.~\cite{Lu2022b}--- opening up opportunities for new regimes of practical Bayesian inference well beyond the experimental four-qubit W state characterized here. And by combining the Bayesian workflow with assumptions such as matrix product forms~\cite{Cramer2010} or low rank~\cite{Gross2010}, as well as recent measurement-efficient approaches like threshold QST~\cite{Binosi2024}, even larger Hilbert spaces can be explored.

We emphasize that our results do not preclude the value of pursuing more orthodox parallelization techniques as well; we are particularly encouraged by recent efforts in coupled chains~\cite{Jacob2020,Heng2023} and parallelized SMC~\cite{Dai2022,Liang2024} that enable parallel implementations avoiding the bias inherent to independent pCN chains. The extent to which these theoretical features can compete with the wall clock speedups from our unorthodox method remains an important open question. In the near-term, therefore, we see our current method not as a replacement for alternative parallel options, but rather an accelerator for existing serial approaches. In light of the observed biases preventing strong parallel scaling [Fig.~\ref{fig:frob}], we believe it would be unwise to fix serial chain length $NT$ \emph{a priori} and then simply scale $R$ in any new problem where the bias plateaus are unknown. Rather, by allocating as many chains $R$ as possible to the problem and increasing individual chains (while monitoring convergence through metrics like the ACF), one can obtain lower errors through parallelization [Fig.~\ref{fig:ACF}(b)] while still relying on the orthodox justification of serial convergence---potentially bringing the best of both serial and parallel worlds to bear in Bayesian QST.

\begin{acknowledgments}
This work was performed in part at Oak Ridge National Laboratory, operated by UT-Battelle for the U.S. Department of energy under contract no. DE-AC05-00OR22725. Funding was provided the U.S. Department of Energy, Office of Science, Advanced Scientific Computing Research (ERKJ432, DE-SC0024257). MATLAB source code for the results in this article is available at \url{https://github.com/hansonnguy/ParallelBayesQST}.
\end{acknowledgments}

\bibliography{references}

\end{document}